\documentclass[11pt,final,onecolumn]{article}
\usepackage{latexsym,amssymb}
\def\RR{{\mathbb R}} \def\CC{{\mathbb C}} \def\QQ{{\mathbb Q}}
\def\PP{{\mathbb P}} \def\ZZ{{\mathbb Z}}

\def\sparc{{\sc Sun Sparc 20}} 
\def\Maple{{\sf Maple}}
\def\Matlab{{\sf Matlab}}			
\def\mixvol{{\sf mixvol}}
\def\spares{{\sf spares}}
\def\far{{\sf far}}

\def\MV{{M \! V}}
\def\vol{{\rm Vol}}
\def\A{{\cal A}}			
\def\B{{\cal B}}
\def\E{{\cal E}}

\newtheorem{theorem}{Theorem}[section]
\newtheorem{definition}[theorem]{Definition}

\begin{document}

\title{\bf Matrix Methods for Solving Algebraic Systems}
\author{Ioannis Z.~Emiris\\
Dept Informatics \& Telecoms, University of Athens, Greece}
\date{March 2014 \footnote{Update of chapter in:
``Symbolic Algebraic Methods and Verification Methods", Springer, 2001.}}
\maketitle 

\begin{abstract}
We present our public-domain software for the following tasks in sparse (or toric) elimination theory, given a well-constrained polynomial system. First, C code for computing the mixed volume of the system. Second, Maple code for defining an overconstrained system and constructing a Sylvester-type matrix of its sparse resultant. Third, C code for a Sylvester-type matrix of the sparse resultant and a superset of all common roots of the initial well-constrained system by computing the eigen-decomposition of a square matrix obtained from the resultant matrix. We conclude with experiments in computing molecular conformations.
\end{abstract}

\section{Introduction}

The problem of computing all common zeros of a system of polynomials
is of fundamental importance in a wide variety of scientific
and engineering applications.
This article surveys efficient methods based on the sparse resultant
for computing all {\em isolated}
solutions of an arbitrary system of $n$ polynomials in $n$ unknowns.
In particular, we construct matrix formulae which yield nontrivial
multiples of the resultant thus reducing
root-finding to the eigendecomposition of a square matrix.

Our methods can exploit structure of the polynomials as well as
that of the resulting matrices.
This is an advantage as compared to other algebraic methods,
such as Gr\"obner bases and characteristic sets.
All approaches have complexity exponential in $n$, but
Gr\"obner bases suffer in the worst case by a quadratic exponent,
whereas for matrix-based methods the exponent is linear.
Moreover, they are discontinuous with respect to perturbations
in the input coefficients, unlike resultant matrix methods in general.
Of course, Gr\"obner bases provide a complete description
of arbitrary algebraic systems and have been well developed,
including public domain stand-alone implementations or as part
of standard computer algebra systems.  
There is also a number of numerical methods for solving algebraic
systems, but their enumeration goes beyond this article's scope.

The next section describes briefly the main steps in
the relatively young theory of sparse elimination,
which aspires to generalize the results of its mature counterpart,
classical elimination.
Section~\ref{Smatrix} presents
the construction of sparse resultant matrices.
Section~\ref{Seigen} reduces solution of arbitrary algebraic systems
to numerical linear algebra, avoiding any issues of convergence.
Our techniques find their natural application in
problems arising in a variety of fields, including problems expressed
in terms of geometric and kinematic constraints in
robotics, vision and computer-aided modelling.
We describe in detail problems from structural biology,
in section~\ref{Sapp_mole}.

The emphasis is placed on recent and new implementations, described in each
respective section, with pointers to where they can be found.
They have been ported on several architectures,
including {\sf Sun, DEC, Linux} and {\sf Iris} platforms.
Previous work and open questions are mentioned in the corresponding
sections.

\section{Sparse elimination} \label{Sspelim}

Sparse elimination generalizes several results of classical elimination
theory on multivariate polynomial systems of arbitrary degree
by considering their structure.
This leads to stronger algebraic and combinatorial results in general
(Gelfand et al.~1994), (Sturmfels 1994), (Cox et al.~1998);
the reader may consult these references for details and proofs.
Assume that the number of variables is $n$;
roots in $(\CC^*)^n$ are called {\em toric}.
We use $x^e$ to denote the monomial $x_1^{e_1} \cdots x_n^{e_n}$,
where $e =(e_1,\ldots,e_n)\in \ZZ^n$.
Let the input {\em Laurent} polynomials be
\begin{equation} \label{square_sys}
f_1,\ldots,f_n\in \QQ[x_1^{\pm 1},\ldots,x_n^{\pm 1}].
\end{equation}
The discussion applies to arbitrary coefficient fields and
roots in the torus of their algebraic closure. 
Let {\em support} $\A_i = \{a_{i1},\ldots,a_{i m_i}\}\subset \ZZ^n$
denote the set of exponent vectors corresponding to monomials in $f_i$
with nonzero coefficients:
$f_i = \sum_{a_{ij}\in\A_i} c_{ij} x^{a_{ij}},$
for $c_{ij}\neq 0.$
The {\em Newton polytope} $Q_i \subset \RR^n$
of $f_i$ is the convex hull of support $\A_i$.
Function {\sf sys\_Maple()} of package \spares\
(see the next section) implements both operations.  
For arbitrary sets $A$ and $B\subset\RR^n$, their
{\em Minkowski sum} is $A+B =$ $\{a+b| \, a\in A, b\in B\}$.

\begin{definition} \label{Din_mv1}
Given convex polytopes $A_1,\ldots,A_n,A_k'\subset\RR^n,$
{\em mixed volume,} is the unique real-valued function
$\MV(A_1,\ldots,A_n)$, invariant under permutations, such that,
$ \MV(A_1,\ldots,\mu A_k + \rho A_k',\ldots,A_n)$
equals
$$
\mu \MV(A_1,\ldots,A_k,\ldots,A_n) + \rho\MV(A_1,\ldots, A_k',\ldots, A_n),
$$
for $\mu,\rho\in\RR_{\geq 0}$,
and, 
$\MV(A_1,\ldots,A_n)=n! \;\vol(A_1)$,
when $A_1=\cdots=A_n$,
where $\vol(\cdot)$ denotes standard euclidean volume in $\RR^n$.
\end{definition}
If the polytopes have integer vertices, their mixed volume takes
integer values.
We are now ready to state a generalization of Bernstein's theorem
(Gelfand et al.~1994), (Cox et al.~1998):

\begin{theorem} \label{Tin_bkk} 
Given system~(\ref{square_sys}), the cardinality of
common isolated zeros in $(\CC^*)^n$, counting multiplicities,
is bounded by $\MV(Q_1,\ldots,Q_n)$.
Equality holds when certain coefficients are generic.
\end{theorem}
Newton polytopes model the polynomials' structure
and provide a ``sparse'' counterpart of total degree.
Similarly for mixed volume and B\'ezout's bound
(simply the product of all total degrees),
the former being usually significantly smaller
for systems encountered in engineering applications.  
The generalization to {\em stable volume} provides
a bound for non-toric roots.  

The algorithm by Emiris and Canny (1995)
has resulted to program \mixvol:\\[10pt]
{\sf Input: supports of $n$ polynomials in $n$ variables}\\
{\sf Output: mixed volume and mixed cells}\\
{\sf Language: C}\\
{\sf Availability: http://www.di.uoa.gr/$\sim$emiris/soft\_geo.html} \\[10pt]
%
Program \mixvol\ enumerates all {\em mixed cells}
in the subdivision of $Q_1+\cdots+Q_n$, thus identifying
the integer points comprising a monomial basis of the quotient ring
of the ideal defined by the input polynomials.
Mixed cells also correspond to start systems (with immediate solution)
for a {\em sparse homotopy} of the original system's roots.
Important work in these areas has been done by T.Y.~Li and his
collaborators (Gao et al.~1999).

The {\em resultant} of a polynomial system of $n+1$ polynomials
in $n$ variables with indeterminate coefficients
is a polynomial in these indeterminates, whose
vanishing provides a necessary and sufficient condition for
the existence of common roots of the system.
Different resultants exist depending on the space of the
roots we wish to characterize, namely projective, affine or toric.
{\em Sparse} or {\em toric resultants} express the existence
of toric roots.
Let
\begin{equation} \label{inp_sys}
f_0,\ldots,f_{n} \in \QQ[x_1^{\pm 1},\dots,x_n^{\pm 1}],
\end{equation}
with $f_i$ corresponding to generic point
$c_i=(c_{i 1}, \ldots, c_{i m_i})$
in the space of polynomials with support $\A_i$.
This space is identified with projective space $\PP^{m_i -1}$.
Then system~(\ref{inp_sys}) can be thought of as point
${c} = (c_0,\ldots,c_{n})$.
Let $Z$ denote the Zariski closure,
in the product of projective spaces,
of the set of all $c$ such that the system
has a solution in $(\CC^*)^n$.
$Z$ is an irreducible variety.

\begin{definition}
The {\em sparse resultant} $R=$ $R(\A_0, \ldots, \A_{n})$
of system (\ref{inp_sys}) is a polynomial in $\ZZ[{c}]$.
If $\mbox{codim} (Z) = 1$ then $R$ is the defining
irreducible polynomial of the hypersurface $Z$.
If $\mbox{codim} (Z) > 1$ then $R= 1$.
\end{definition}
The resultant is homogeneous in the coefficients of each polynomial.
If $\MV_{-i} =$
$\MV(Q_0, \ldots, Q_{i-1}, Q_{i+1}, \ldots, Q_{n}),$
then the degree of $R$ in the coefficients of
$f_i$ is $\deg_{f_i} R = \MV_{-i}$.
$\deg R$ will stand for the total degree.

\section{Matrix formulae} \label{Smatrix}

Different means of expressing a resultant are possible
(Cox et al.~1998), (Emiris and Mourrain 1999).
Ideally, we wish to express it as a matrix determinant, or a
divisor of such a determinant where the quotient is a
nontrivial extraneous factor.
This section discusses matrix formulae for the sparse resultant,
which exploit the monomial structure of the Newton polytopes.
These are {\em sparse resultant}, or {\em Newton, matrices}.
We restrict ourselves to Sylvester-type matrices which generalize
the coefficient matrix of a linear system and Sylvester's matrix
of two univariate equations.

There are two main approaches to construct a
well-defined, square, generically nonsingular matrix $M$, 
such that $R | \det M$.
The rows of $M$ will always be indexed by the product
of a monomial with an input polynomial.
The entries of a row are coefficients of that product, each
corresponding to the monomial indexing the respective column.
The degree of $\det M$ in the coefficients of $f_i$,
equal to the number of rows with coefficients of $f_i$,
is greater or equal to $\deg_{f_i} R$.  
Obviously, the smallest possible matrix has dimension $\deg R$.

The first approach, introduced by Canny and Emiris in~1993,
relies on a {\em mixed subdivision} of the
Minkowski sum of the Newton polytopes $Q =$ $Q_0 + \cdots + Q_{n}$
(Canny and Pedersen~1993), (Sturmfels~1994), (Canny and Emiris~2000).
The algorithm uses a subset of $(Q+\delta) \cap \ZZ^n$
to index the rows and columns of $M$.
$\delta\in\QQ^n$ must be {\em sufficiently generic} so that
every integer point lies in the relative interior of a unique
$n$-dimensional cell of the mixed subdivision of $Q+\delta$.
In addition, $\delta$ is small enough so that this cell is among
those that had the point on their boundary.
Clearly, the dimension of the resulting matrix is at most equal
to the number of points in $(Q+\delta) \cap \ZZ^n$.
This construction allows us to pick any one polynomial so that
it corresponds to exactly $\deg_{f_i} R$ rows.

The greedy version of Canny and Pedersen~(1993)
uses a minimal point set
and is the algorithm implemented by function {\sf spares()}
in the \Maple\  package of the same name.
It is also included as function {\sf spresultant()} in \Maple\  package
{\sf multires} developed at INRIA
({\sf http://www-sop.inria.fr/galaad/logiciels/multires.html}):\\[10pt]
{\sf Input: $n+1$ polynomials in $n$ variables,
	an arbitrary number of parameters}\\
{\sf Output: sparse resultant matrix in the parameters}\\
{\sf Language: \Maple}\\
{\sf Availability: http://www.di.uoa.gr/$\sim$emiris/soft\_alg.html}\\[10pt]
For instance, {\sf spares([f0,f1,f2],[x1,x2])}
constructs the sparse resultant matrix of the 3 polynomials
by eliminating variables {\sf x1, x2}.
The function also expresses the polynomial coefficients
in terms of any indeterminates other than {\sf x1, x2}.
Optional arguments may specify vector $\delta$ and
the subdivision of $Q$.

The second algorithm, by Emiris and Canny (1995),
is {\em incremental} and yields usually smaller matrices and, in any case,
of dimension no larger than the cardinality of $(Q+\delta) \cap \ZZ^n$.
We have observed that in most cases of systems with dimension bounded
by 10 the algorithm gives a matrix at most 4 times the optimal.
The selection of integer points corresponding to monomials multiplying
the row polynomials uses a vector $v\in(\QQ^*)^n$.
In those cases where a minimum matrix of Sylvester type provably exists,
the incremental algorithm produces this matrix.
These are precisely the
systems for which $v$ can be deterministically specified;
otherwise, a random $v$ can be used.  

The algorithm proceeds by constructing candidate rectangular matrices
in the input coefficients.
Given such a matrix with the coefficients specialized to generic values,
the algorithm verifies whether its rank is complete
using modular arithmetic.
If so, any square nonsingular submatrix can be returned as $M$;
otherwise, new rows (and columns) are added to the candidate.
This is the first part of program \far.
The entire \far\  has:\\[10pt]
{\sf Input: $n+1$ polynomials in $n$ variables to be eliminated,
one in the coefficient field}\\
{\sf Output: sparse resultant matrix and a superset of the common roots}\\
{\sf Language: C}\\
{\sf Availability: http://www.di.uoa.gr/$\sim$emiris/soft\_alg.html}\\[10pt]
For instance, commands {\sf ``far -nco trial input''}
and {\sf ``far -nco -ms 0 trial input''} construct a sparse resultant matrix,
where file {\sf input.exps} contains the supports
and file {\sf input.coef} contains vector $v$ and the $\MV_{-i}$,
if known (otherwise the program computes them by calling \mixvol\
and writes them in file {\sf temp\_all\_mvs}).
In the first case, we assumed file {\sf trial.msum} exists and contains all
needed integer points for matrix construction.
In the second example this file is created and filled in by \far.
A number of command line options exists, including {\sf ``-iw trial.indx''}
to store the matrix definition in file {\sf trial.indx} in order to be
used by subsequent executions.

Sparse resultant matrices, including the candidates constructed
by the incremental algorithm, are characterized by a structure that
generalizes the Toeplitz structure and has been called
{\em quasi-Toeplitz} (Emiris and Pan 2002).
An open implementation problem is to exploit this structure in
verifying full rank, aspiring to match
the asymptotic acceleration of almost one order of magnitude.
Another open question concerns exploiting quasi-Toeplitz structure
for accelerating the solution of an eigenproblem.

D'Andrea (2002) proved that, if the mixed subdivision is constructed
carefully, it is possible to obtain Macaulay-type formulae for the sparse
resultant, namely to define a submatrix whose determinant yields the
extraneous factor.
Emiris and Konaxis (2011) simplified this construction with the aim
of obtaining an implementation.

\section{Algebraic solving by linear algebra} \label{Seigen}

To solve the well-constrained system~(\ref{square_sys})
by the resultant method we define an overconstrained
system and apply the resultant matrix construction.
Matrix $M$ need only be
computed once for all systems with the same supports.
So this step can be carried out offline, while the
matrix operations to approximate all isolated roots
for each coefficient specialization are online.

We present two ways of defining an overconstrained system.
The first method
{\em adds an extra polynomial} $f_0$ to the given system
(thus defining a well-studied object, the $u$-resultant).
The constant term is a new indeterminate:
$$
f_0 = x_0 + c_{01} x_1 + \cdots + c_{0n} x_n\;\in
(\QQ[x_0])[x_1^{\pm 1},\dots,x_n^{\pm 1}].
$$
Coefficients $c_{0j}$ are usually random.
$M$ describes the multiplication map
for $f_0$ in the coordinate ring of the ideal defined by~(\ref{square_sys}).
An alternative way to obtain an overconstrained system 
is by {\em hiding} one of the variables in the coefficient field
and consider
(after modifying notation to unify the subsequent discussion)
system:
$$
f_0,\ldots,f_{n} \in
\left(\QQ[x_{0}]\right) [x_1^{\pm 1},\ldots,x_{n}^{\pm 1}].
$$
$M$ is a matrix polynomial in $x_0$, 
and may not be linear.  

In both cases, the idea is that when $x_0$ is equal to the respective
coordinate of a common root, then the resultant and, hence, the matrix
determinant vanish.
An important issue
concerns the degeneracy of the input coefficients.
This may result in the trivial vanishing of the sparse resultant
or of $\det M$ when there is an infinite number of common roots
(in the torus or at toric infinity) or 
simply due to the matrix constructed.
An infinitesimal perturbation has recently been proposed
by D'Andrea and Emiris (2001),
which respects the structure of Newton
polytopes and is computed at minimal extra cost.

The perturbed determinant is a polynomial in the perturbation variable,
whose leading coefficient is guaranteed to be nonzero.
The trailing
{\em nonzero} coefficient is always a multiple of a generalized resultant,
in the sense that it vanishes when $x_0$ takes its values at
the system's roots.
This is a univariate polynomial in $x_0$, hence
univariate equation solving yields these coordinates.
Moreover, the $u$-resultant allows us to recover all coordinates
via polynomial factorization.
The perturbed matrix can be obtained by package \spares,
provided that local variable {\sf PERT\_DEGEN\_COEFS} is
appropriately set, as explained in the package's documentation.
An open problem concerns the combination of this perturbation with
the matrix operations described below.

\subsection{Eigenproblems} \label{Spo_eigen}

This section describes the online matrix solver of \far.
Most of the computation is numeric, yet the method has global convergence
and avoids issues related to the starting point of iterative methods.
We use double precision floating point arithmetic and
the LAPACK library because it implements state-of-the-art
algorithms, offering the choice of a tradeoff between
speed and accuracy, and provides efficient ways for computing
estimates on the condition numbers and error bounds.

A basic property of resultant matrices is that right vector
multiplication expresses evaluation of the row polynomials.
Specifically, multiplying by a column vector
containing the values of column monomials $q$ at some
$\alpha\in\CC^{n}$ produces the values of the row
polynomials $\alpha^p f_{i_p}(\alpha)$, where integer point (or,
equivalently, monomial) $p$ indexes a row.
Letting $\E$ be the monomial set indexing the matrix rows and columns,
$$
M(x_0)
\left[ \begin{array}{c} \vdots\\ \alpha^q\\ \vdots\\ \end{array} \right]
= \left[ \begin{array}{c} \vdots\\
\alpha^p f_{i_p}(x_0,\alpha)\\ \vdots\\ \end{array} \right],
\quad q,p\in\E, i_p\in\{0,\ldots,n\}.
$$
Computationally it is preferable to have to deal with as small a matrix as
possible.
To this end we partition $M$ into four blocks $M_{ij}$
so that the upper left submatrix $M_{11}$ is of maximal
possible dimension under the following conditions:
it must be square, independent of $x_{0}$, and well-conditioned
relative to some user-defined threshold.
\far\ first concentrates all constant columns to the left and within
these columns permutes all zero rows to the bottom;
both operations could be implemented offline.
To specify $M_{11}$ according to the above conditions,
an LU decomposition with column pivoting is applied,
though an SVD (or QR decomposition) might be preferable.

Once $M_{11}$ is specified, let $A(x_{0}) =$
$M_{22}(x_{0}) - M_{21}(x_{0}) M_{11}^{-1} M_{12}(x_{0})$.
To avoid computing $M_{11}^{-1}$, we use its
decomposition to solve $M_{11} X = M_{12}$ and
compute $A = M_{22} - M_{21} X$.
The routine used depends on $\kappa(M_{11})$, with
the slower but more accurate function {\sf dgesvx} 
called when $\kappa(M_{11})$ is beyond some threshold.

If $(\alpha_0,\alpha)\in\CC^{n+1}$ is a common root with
$\alpha\in\CC^n$, then $\det M(\alpha_{0})=0 \Rightarrow$
$\det A(\alpha_{0}) = 0$.
Let point (or monomial) set $\B\subset\E$ index matrix $A$.  
For any vector $v'=[\cdots\alpha^{q}\cdots]^T$,
where $q$ ranges over $\B$, $A(\alpha_{0}) v'=0$.
Moreover,
\begin{eqnarray*}
\left[ \begin{array}{cc} M_{1 1} & M_{1 2}(\alpha_{0}) \\
			 0  & A(\alpha_{0})   \\ \end{array} \right]
\left[ \begin{array}{c} v\\ v'\\ \end{array} \right]
= \left[ \begin{array}{c} 0\\ 0\\ \end{array} \right] & \Rightarrow &
M_{11} v + M_{12}(\alpha_{0}) v' = 0,
\end{eqnarray*}
determines $v$ once $v'$ has been computed.  
Vectors  $v$ and $v'$ together contain the values of every
monomial in $\E$ evaluated at $\alpha$.

It can be shown that $\E$ affinely spans $\ZZ^n$ and
an affinely independent subset can be computed in
polynomial time.
Given $v$, $v'$ and these points,
we can compute the coordinates of $\alpha$.
If all independent points are in $\B$ then $v'$ suffices.
To find the vector entries that will allow us to recover the
root coordinates it is sufficient to search
for pairs of entries corresponding to $q_1, q_2$
such that $q_1-q_2=(0,\ldots,0,1,0,\ldots,0)$.
This lets us compute the $i$-th coordinate, if the unit appears
at the $i$-th position, by taking ratios of the vector entries.

To reduce the problem to an eigendecomposition,
let $r$ be the dimension of $A(x_0)$,
and $d\geq 1$ the highest degree of $x_0$ in any entry.
We wish to find $x_0:$
$$
A(x_0)=x_0^d A_d + x_0^{d-1} A_{d-1} + \cdots + x_0 A_1 + A_0
$$
becomes singular.
These are the eigenvalues of the {\em matrix polynomial}.
Furthermore, for every eigenvalue $\lambda$, there is a basis
of the kernel of $A(\lambda)$ defined by the {\em right eigenvectors}
of the matrix polynomial.
If $A_d$ is nonsingular then the eigenvalues and right eigenvectors of $A(x_0)$
are the eigenvalues and right eigenvectors of {\em monic} matrix polynomial
$A_d^{-1} A(x_0)$.
This is always the case when adding an extra linear polynomial,
since $d=1$ and $A_1=I$ is the $r\times r$ identity matrix;
then $A(x_0)=-A_1(-A_1^{-1}A_0-x_0I)$.
Generally, the {\em companion matrix} of a monic matrix polynomial is
a square matrix $C$ of dimension $rd$:
$$
C = \left[ \begin{array}{cccc}
	0       & I     & \cdots        & 0     \\
	\vdots  &       & \ddots        &       \\
	0       & 0     & \cdots        & I     \\
	-A_d^{-1}A_0    & -A_d^{-1}A_1 & \cdots & -A_d^{-1}A_{d-1} \\
	\end{array} \right].
$$
The eigenvalues of $C$ are precisely the eigenvalues $\lambda$ of 
$A_d^{-1} A(x_0)$, whereas its right eigenvector equals the
concatenation of $v_1,\ldots,v_d :$
$v_1$ is a right eigenvector of $A_d^{-1} A(x_0)$ and
$v_i=\lambda^{i-1} v_1$, for $i=2,\ldots,d$.
There is an iterative and a direct algorithm in LAPACK for solving this
eigenproblem, respectively implemented in {\sf hsein} and {\sf trevc}.
Experimental evidence points to the former as being faster on 
large problems.
Further, an iterative solver could eventually exploit the fact
that we are only interested in real eigenvalues and eigenvectors.

We now address the question of a singular $A_d$.
The following {\em rank balancing} transformation is used in \far\
also to improve the conditioning of the leading matrix:
If matrix polynomial $A(x_0)$ is not singular for all $x_0$,
then there exists a transformation
$x_0\mapsto (t_1 y + t_2)/(t_3 y + t_4)$ for some
$t_i\in\ZZ$, that produces a new matrix polynomial
of the same degree and with
nonsingular leading coefficient.  
The new matrix polynomial has coefficients
of the same rank, for sufficiently generic $t_i$.
We have observed that for matrices of dimension larger than 200,
at least two or three quadruples should be tried since a lower condition
number by two or three orders of magnitude is sometimes achieved.
The asymptotic as well as practical
complexity of this stage is dominated by the eigendecomposition.

If a matrix polynomial with invertible leading matrix is found, then the
eigendecomposition of the corresponding companion matrix is undertaken.
If $A_d$ is ill-conditioned for all linear rank balancing transformations,
then we build the matrix pencil and call the
{\em generalized eigendecomposition} routine {\sf dgegv} to solve $C_1x + C_0$.
The latter returns pairs $(\alpha,\beta)$ such that matrix
$C_1 \alpha + C_0 \beta$ is singular with an associated
right eigenvector.
For $\beta\neq 0$ the eigenvalue is $\alpha/\beta$, while for
$\beta=0$ we may or may not wish to discard the eigenvalue.
$\alpha=\beta=0$ occurs if and only if the pencil is identically zero
within machine precision.

If the $x_0$-root coordinates have all unit {\em geometric multiplicity}
and $A(x_0)$ is not identically singular, then we
have reduced root-finding to an eigenproblem and some evaluations
to eliminate extraneous eigenvectors and eigenvalues.
The complexity lies in
$O^*\left( 2^{O(n)} (\deg R)^3 d \right),$
where polylogarithmic terms are ignored.

These operations are all implemented in the solver of \far:
Command {\sf ``far trial input''}
constructs the sparse resultant matrix and solves the system with
supports in {\sf input.exps},
vector $v$, mixed volumes $\MV_{-i}$ and (integer) coefficients in
{\sf input.coef}, and integer points in {\sf trial.msum}.
Command line option {\sf ``-ir trial.indx''} reads the matrix
definition from file {\sf trial.indx} and option {\sf -a}
tells the program to set $A(x_0)=M(x_0)$, thus avoiding the
decomposition of $M_{11}$ and any related numerical errors.
Other options control the condition number thresholds,
printing of various information, and verification of results.

The section concludes with accuracy issues, irrespective of
whether $A_d$ is regular or not.
Since there is no restriction in picking which variable to hide, it is enough
that one of the original $n+1$ variables have unit geometric multiplicity.
If none can be found, we can specialize the hidden variable to each of the
eigenvalues and solve every one of the resulting subsystems.  
Other numerical and algebraic remedies are under study, including
the aforementioned perturbed determinant.
Still, there is an open implementation problem in
verifying the multiplicity and solving in such cases.
Clustering neighbouring eigenvalues and computing the error on the
average value will help handling such cases.
Lastly, self-validating methods should be considered to handle
ill-conditioned matrix polynomials, in particular in the presence of
defective eigenvalues.

\section{Molecular conformations} \label{Sapp_mole}

A relatively new branch of computational biology has been emerging as an effort
to apply successful paradigms and techniques from geometry and
robot kinematics to predicting the structure
of molecules and embedding them in euclidean space.
This section examines the problem of computing all
3-dimensional {\em conformations} of a
molecule described by certain geometric characteristics.

Energy minima can
be approximated by allowing only the dihedral angles to vary, while 
considering bond lengths and bond angles as rigid.
We consider cyclic molecules of six atoms to illustrate our approach
and show that the corresponding algebraic formulation conforms to our model
of sparseness.
An in-depth study of cyclic molecules has been presented
by Emiris and Mourrain (1999).
Direct geometric analysis
yields a $3\times 3$ polynomial system
$$
f_i = \beta_{i1} + \beta_{i2} x_j^2 + \beta_{i3} x_k^2 +
	\beta_{i4} x_j^2 x_k^2 + \beta_{i5} x_j x_k = 0,
\quad i\in\{1,2,3\},
$$
for $\{i,j,k\}=\{1,2,3\}$. 
The $\beta_{ij}$ are functions of known parameters.
The system has a B\'ezout bound of 64 and mixed volume 16.

The first instance tried is a synthetic example for which
$\beta_{ij}$ is the $j$-th entry of $(-9,-1,-1,3,8)$
for all $i$.
The symmetry of the problem is bound to produce root
coordinates of high multiplicity, so we add
$f_0 =$ $x_0 + c_{01} x_1 - c_{02} x_2 + c_{03} x_3$
with randomly selected $c_{0j}$.
The 3-fold mixed volumes are $16, 12,12,12$ hence $\deg R=52$.
$M$ is regular and
has dimension 86, with 30 rows corresponding to $f_0$.
The entire $56\times 56$ constant submatrix is relatively
well-conditioned.
In the $30\times 30$ matrix polynomial,
matrix $A_1$ is numerically singular;
random transformations fail to improve significantly
its conditioning.
The generalized eigenproblem routine
produces 12 complex solutions, 3 infinite real solutions
and 15 finite real roots.
The absolute value of the four polynomials on the
candidate values lies in $[0.6\cdot 10^{-9}, 0.3\cdot 10^{-3}]$
for values that approximate true solutions and in
$[7.0,3.0\cdot 10^{20}]$ for spurious answers.
Our program computes the true roots to at least 5 digits,
the true roots being
$\pm(1,1,1),\, \pm(5,-1,-1),$ $\pm(-1,5,-1),\, \pm(-1,-1,5).$
The average CPU time of the online phase on a \sparc\ 
with clock rate 60MHz and 32MB of memory
is $0.4$ seconds.

Usually noise enters in the process that produces the coefficients.
To model this phenomenon,
we consider the {\em cyclohexane} which has
equal inter-atomic distances and equal bond angles.
We randomly perturb these values by about $10\%$ to obtain
$$
\beta = \left[ \begin{array}{rrrrr}
	-310& 959 & 774& 1313& 1389\\
	-365& 755 & 917& 1269& 1451\\
	-413& 837 & 838& 1352& 1655
\end{array} \right].
$$
We defined an overconstrained system by hiding variable $x_3$.
$M$ has dimension 16 and is quadratic in $x_3$,
whereas the 2-fold mixed volumes are all 4 and $\deg R=12$.  
The monic quadratic polynomial reduces to a $32\times 32$ companion matrix
on which the standard eigendecomposition is applied.
After rejecting false candidates
each solution contains at least 8 correct digits.
CPU time is $0.2$ seconds on average for the
online phase.

Last is an instance where the input parameters are
sufficiently generic to produce 16 real roots.
Let $\beta_{ij}$ be the $j$-th entry of $(-13, -1, -1, -1, 24)$.
We hide $x_3$ and obtain $\dim M= 16$,
whereas $\deg R=12$, and the companion matrix has dimension 32.
There are 16 real roots.
Four of them correspond to eigenvalues of unit geometric multiplicity,
while the rest form four groups, each corresponding to a triple eigenvalue.
For the latter the eigenvectors give us no valid information, so we
recover the values of $x_1,x_2$ by looking at the other solutions and
by relying on symmetry.
The computed roots are correct to at least 7 decimal digits.
The average CPU time is $0.2$ seconds.

An equivalent approach to obtaining the same algebraic system
may be based on {\em distance geometry}.
A {\em distance matrix} is a square,
real symmetric matrix, with zero diagonal.
It can encode all inter-atomic distances by associating its
rows and columns to atoms.
When the entries are equal to a scalar multiple
of the corresponding squared pairwise distance, the matrix
is said to be {\em embeddable} in $\RR^3$.
Necessary and sufficient conditions for such matrices to be
embeddable are known in terms of the
eigenvalues and rank.

The main interest of this approach lies in large molecules.
We have examined it in relation with experimental data that
determine intervals in which the unknown distances lie.
Optimization methods have been developed and applied successfully
to molecules with a few hundreds of atoms (Havel et al.~1997).
Ours are direct linear algebra techniques which are, for now,
in a preliminary stage.
We apply results from distance matrix theory and
structured matrix perturbations to reduce
the rank of the interval matrix respecting the experimental
bounds.
The \Matlab\  code developed by
Emiris and Nikitopoulos (2005) can handle molecules with up to $30$ atoms.

\section*{References}
{\small
\noindent
Canny, J., Emiris, I.Z.~(2000):
A subdivision-based algorithm for the sparse resultant.
J.~ACM.  47:417--451\\[-4pt]

\noindent
Canny, J., Pedersen, P. (1993):
An algorithm for the {N}ewton resultant.
Technical Report 1394, Computer Science Department,
Cornell University, Ithaca, New York\\[-4pt]

\noindent
Cox, D., Little, J., O'Shea, D. (1998):
Using algebraic geometry.
Springer, New York (Graduate Texts in Mathematics, vol.~185)\\[-4pt]

\noindent
D'Andrea, C. (2002):
Macaulay-style formulas for the sparse resultant.
Trans.\ of the AMS, 354:2595--2629\\[-4pt]

\noindent
D'Andrea, C., Emiris, I.Z. (2001):
Computing Sparse projection operators.
In: Symbolic Computation: Solving Equations in Algebra, Geometry,
and Engineering, eds Green E.L.\ et al.
Contemporary Mathematics, vol.~286, pages 121--139, AMS
\\[-4pt]

\noindent
Emiris, {I.Z.}, Canny, {J.F.} (1995):
Efficient incremental algorithms for the sparse resultant and the
  mixed volume.
J.~Symb.\  Comput. 20:117--149\\[-4pt]

\noindent
Emiris, {I.Z.}, Konaxis, C. (2011):
Single-lifting {Macaulay-type} formulae of generalized unmixed sparse
resultants.
J.\ Symb.\ Computation, 46(8):919--942\\[-4pt]

\noindent
Emiris, {I.Z.}, Mourrain, B. (1999):
Computer algebra methods for studying and computing molecular
 conformations.
Algorithmica, Special issue on algorithms for computational
 biology. 25:372--402\\[-4pt]

\noindent
Emiris, I.Z., Nikitopoulos, T.G.~(2005):
Molecular Conformation Search by distance matrix perturbations.
J.\ Math.\ Chemistry, 37(3):233--253 \\[-4pt]

\noindent
Emiris, {I.Z.}, Pan, V.Y.\  (2002):
Symbolic and Numeric Methods for exploiting structure in 
constructing resultant matrices.
J.\ Symb.\ Computation, 33:393--413\\[-4pt]

\noindent
Gao, T., Li, T.Y., Wang, X. (1999):
Finding isolated zeros of polynomial systems in {${C}^n$}
with stable mixed volumes.  J.~Symb.\  Comput.  
28:187--211\\[-4pt]

\noindent
Gelfand, I.M., Kapranov, M.M., Zelevinsky, A.V.~(1994):
Discriminants and resultants.
Birkh\"{a}user, Boston\\[-4pt]

\noindent
Havel, T.F., Hyberts, S., Najfeld, I.~(1997):
Recent advances in molecular distance geometry.
In: Bioinformatics, Springer, Berlin
(Lecture Notes in Computer Science, vol.~1278)\\[-4pt]

\noindent
Sturmfels, B.~(1994):
On the {N}ewton polytope of the resultant.
J.~Algebr.\  Combinat. 3:207--236
}
\end{document}